\documentclass[12pt]{article}
\usepackage{a4wide}
\usepackage{amssymb}
\usepackage{amsmath}
\usepackage{graphicx}
\usepackage{mathdots}
\usepackage{slashed}
\usepackage{verbatim}
\usepackage{xcolor}
\usepackage[utf8]{inputenc}

\usepackage{hyperref}

\usepackage{IEEEtrantools}

\def\makeatletter{\catcode`\@=11}% 11:letter
\makeatletter
\def\mathbox#1{\hbox{$\m@th#1$}}%
\def\math@ccstyles#1#2#3#4#5#6#7{{\leavevmode
      \setbox0\mathbox{#6#7}%
      \setbox2\mathbox{#4#5}%
      \dimen@ #3%
      \baselineskip\z@\lineskiplimit#1\lineskip\z@
      \vbox{\ialign{##\crcr
             \hfil \kern #2\box2 \hfil\crcr
             \noalign{\kern\dimen@}%
             \hfil\box0\hfil\crcr}}}}
\def\mathaccstyles{\math@ccstyles\maxdimen}
\def\maththroughstyles{\math@ccstyles{-\maxdimen}}
\def\unity%
 {\maththroughstyles{.45\ht0}\z@\displaystyle {\mathchar"006C}\displaystyle 1}
 
\usepackage{tikz}
\usetikzlibrary{positioning,arrows}
\usetikzlibrary{decorations.pathmorphing,calc}
\usetikzlibrary{decorations.markings}
\newif\ifmirrorsemicircle
\usepackage{float}
\usepackage{ifthen}
\usepackage{enumerate}
\usepackage{amsmath}
\usepackage{amssymb}
\usepackage[mathscr]{eucal}
\usepackage[errorshow]{tracefnt}
\usepackage{amsmath}
\usepackage{slashed}
\usepackage{amssymb}
\usepackage{array}
\usepackage{amsthm}
\usepackage{eucal}
\usepackage{epsf}
\usepackage{epsfig}
\usepackage{psfrag}
\usepackage{multirow}
\usepackage{wasysym} 
\usepackage{tikz-cd} 
\usepgfmodule{shapes}
\usetikzlibrary{backgrounds, arrows,calc,shapes,decorations.pathreplacing, decorations.markings, automata,positioning}
\usepackage{verbatim}

\begin{document}

\begin{flushright}\footnotesize

\texttt{ICCUB-19-018}
\vspace{0.6cm}
\end{flushright}

\mbox{}
\vspace{0truecm}
\linespread{1.1}

%%%%%%%%%%%%%%%%%
\centerline{\LARGE \bf Correlation functions in scalar field theory}
\medskip

\centerline{\LARGE \bf  at large charge}

\medskip

%\centerline{\LARGE \bf } 

\vspace{.4cm}

 \centerline{\LARGE \bf }

\vspace{1.5truecm}

\centerline{
    {\bf G. Arias-Tamargo${}^{a,b}$} \footnote{ariasguillermo@uniovi.es},
    { \bf D. Rodriguez-Gomez${}^{a,b}$} \footnote{d.rodriguez.gomez@uniovi.es},
%    {\bf and}
    { \bf J. G. Russo ${}^{c,d}$} \footnote{jorge.russo@icrea.cat}}

\vspace{1cm}
\centerline{{\it ${}^a$ Department of Physics, Universidad de Oviedo}} \centerline{{\it C/ Federico Garc\'ia Lorca  18, 33007  Oviedo, Spain}}
\medskip
\centerline{{\it ${}^b$  Instituto Universitario de Ciencias y Tecnolog\'ias Espaciales de Asturias (ICTEA)}}\centerline{{\it C/~de la Independencia 13, 33004 Oviedo, Spain.}}
\medskip
\centerline{{\it ${}^c$ Instituci\'o Catalana de Recerca i Estudis Avan\c{c}ats (ICREA)}} \centerline{{\it Pg.~Lluis Companys, 23, 08010 Barcelona, Spain}}
\medskip
\centerline{{\it ${}^d$ Departament de F\' \i sica Cu\' antica i Astrof\'\i sica and Institut de Ci\`encies del Cosmos}} \centerline{{\it Universitat de Barcelona, Mart\'i Franqu\`es, 1, 08028
Barcelona, Spain }}
\vspace{1cm}

\centerline{\bf ABSTRACT}
\medskip 
We  compute general higher-point functions in the sector of large charge operators $\phi^n$, $\bar\phi^n$ at large charge
in  $O(2)$ $(\bar \phi\phi)^2$ theory.
We find that there is a special class of ``extremal" correlators having only one insertion of  $\bar \phi^n$
that have a remarkably simple form in the double-scaling limit $n\to \infty $ at fixed $g\,n^2\equiv \lambda$,
where $g\sim\epsilon $ is the coupling at the $O(2)$ Wilson-Fisher fixed point in $4-\epsilon$ dimensions. 
In this limit, also  non-extremal correlators can be computed. As an example, we give the complete formula for
$ \langle \phi(x_1)^{n}\,\phi(x_2)^{n}\,\bar{\phi}(x_3)^{n}\,\bar{\phi}(x_4)^{n}\rangle$, which reveals an  interesting structure.

\noindent

\newpage

%\tableofcontents

\section{Introduction}

 Recently it has been appreciated that sectors with large charge under a global symmetry of a given quantum field theory enjoy remarkable simplification properties, which allow a systematic and analytic study \cite{Hellerman:2015nra,Alvarez-Gaume:2016vff,Monin:2016jmo}. A prototypical example is the $O(2)$ model in $d=3$, where, if one is interested in the sector of operators with large charge $n$ under the global symmetry, it is possible to write an effective field theory governing their dynamics. This allows to compute  their anomalous dimensions, which are found to scale as $\Delta\sim n^{\frac{3}{2}}+\mathcal{O}(n^{\frac{1}{2}})$. This result can be understood from a ``microscopic" description starting with the $U(1)$ Wilson-Fisher (WF) fixed point in $d=4-\epsilon$ dimensions, described by the action  

\begin{equation}
S_0 =\int d^{4-\epsilon}x\,\left(\partial\bar \phi\,\partial\phi-\frac{g}{4}\,(\bar\phi \phi)^2\right)\,.
\end{equation}
This approach in fact uncovers a  rich structure, as the sector of operators at fixed charge $n$ is described by an effective theory depending on both $n$ and $g$ (recall that at the WF fixed point, $g\sim \epsilon$), in such a way that, depending on how the large charge limit is taken, a different behavior emerges. 
Some aspects of the large $n$ expansion were recently investigated in
 \cite{Arias-Tamargo:2019xld,Badel:2019oxl,Watanabe:2019pdh,Badel:2019khk},  where the two-point function  of the operators $\phi^n$, $\bar\phi^n $  was computed at large $n$. 
 In particular, a general way to organize the expansion was discussed in 
  \cite{Badel:2019oxl,Watanabe:2019pdh,Badel:2019khk}
 (generalizing earlier work in  \cite{Libanov:1994ug,Libanov:1995gh,Son:1995wz}).
  The 
 effective description of the large $n$ sector naturally depends on $n$ and $g\,n=\hat \lambda$, in such a way that
  a 't Hooft-like double expansion emerges. For the dimension of the operator $\phi^n$ one has
\begin{equation}
\label{2point}
\Delta= \sum_{k=-1}^\infty \,n^{-k}\ \Delta_k(\hat \lambda )\ .
\end{equation}
In the strong coupling regime --which overlaps with the regime of validity of the large charge effective theory--, one finds $\Delta\sim \hat\lambda^{\frac{4-\epsilon}{3-\epsilon}}+\cdots$, thus recovering the expected scaling $\Delta \sim n^{\frac{d}{d-1}}$, with the extra bonus of providing an analytic expression for the actual coefficients in the large charge expansion \cite{Badel:2019oxl,Watanabe:2019pdh,Badel:2019khk}.  

On the other hand, at weak coupling and in the 
%strict 
large $n$ limit, the dominant term is $\Delta_{-1}(\hat \lambda)$, which must admit a  perturbative expansion for small $\hat\lambda$. This gives
\cite{Arias-Tamargo:2019xld,Badel:2019oxl,Watanabe:2019pdh}
\begin{equation}
\Delta_{-1}(\hat\lambda ) =  1 + \frac{\hat \lambda}{32\pi^2} + O\big(\hat\lambda^2 \big)\,.
\end{equation}
In \cite{Arias-Tamargo:2019xld}, the leading correction
was obtained by a full resummation of the dominant Feynman diagrams  --dubbed Kermit $L$-loop diagrams-- that survive in 
a double scaling limit with $n\rightarrow \infty$ at fixed $g\,n^2\equiv \lambda$.
In this limit the sum over Feynman diagrams exponentiates, giving the result
\begin{equation}
    \langle \phi^n(x) \bar\phi^n(0)\rangle = \frac{n!}{(4\pi^2)^n |x|^{2\Delta} }\ ,\qquad \Delta = n+\frac{\lambda}{32\pi^2}\,.
\label{ddelta}
\end{equation}
This result can also be derived by a saddle-point evaluation of the two-point function, which becomes
exact in the double scaling limit with fixed $\lambda =g n^2 $
\cite{Arias-Tamargo:2019xld}.

In the double-scaling limit of \cite{Arias-Tamargo:2019xld} at fixed $\lambda $, the  
 $O\big(\hat\lambda^2 \big)$ terms are given by Feynman diagrams which are suppressed by powers of $1/n$.
Thus, from this point of view, the result \eqref{ddelta}
can be viewed as the leading term of the more general double expansion in $n$, $\hat \lambda$. 
 
Large charge expansions also exist
in general  CFT's with a marginal coupling. An example of a CFT depending on an exactly marginal parameter $g_{\rm YM}$ is   $\mathcal{N}=2$ supersymmetric four-dimensional QCD with  gauge group $SU(N)$ and $2N$ fundamental flavors.
The large charge limit of this theory was first introduced in \cite{Bourget:2018obm} and studied using supersymmetric localization. Localization leads to exact formulas for
a special class of correlation functions of superconformal chiral primary operators, called ``extremal correlators".
These are correlation functions with  an arbitrary number of
insertions of operators ${\rm Tr}\, \phi^n$ and only one insertion of ${\rm Tr}\, \bar \phi^n$ and they enjoy special properties because of supersymmetry ($\phi$ being  the adjoint scalar in the vector multiplet of unit R-charge).
It was shown that the perturbative expansion of correlators of $({\rm Tr}\, \phi^2)^n$ has a well-defined large $n$ limit provided one takes 
a double-scaling limit  of large $n$  and fixed $g_{\rm YM}^2\,n$. This limit ensures that all terms in the perturbative expansion are
finite and non-vanishing. Further aspects were studied in detail in \cite{Beccaria:2018xxl,Beccaria:2018owt}. Subsequently, the existence of a double-scaling limit was understood in terms of a ``hidden" matrix model description in \cite{Grassi:2019txd}.

The mere fact that it is possible to  compute observables of a QFT in a closed form  in the large charge sector is remarkable \textit{per se}. Motivated by this, in this paper, we study higher point functions in the $O(2)$ theory in the sector of operators with large charge. Focusing in the weak coupling regime in the double expansion in $1/n$, $\hat\lambda$, we  compute ``extremal" correlators (of the form $\langle \phi^{n_1}\cdots\phi^{n_r}\bar\phi^m\rangle$) as well as 4-point functions in the ``non-extremal" case. As discussed above, in the double scaling limit,  these results become exact. We shall use the saddle point method employed in \cite{Arias-Tamargo:2019xld}.

\section{Higher point functions in the $O(2)$ model}

We will follow the approach of \cite{Arias-Tamargo:2019xld}, where the two-point function was computed in a double-scaling limit,  $n\to \infty $, $g\to 0$ at fixed $g\,n^2\equiv \lambda$.
This limit yields the exact exponentiation of the
 the leading non-trivial term in the more general $n$, $\hat\lambda$ expansion in the large $n$ and weak $\hat\lambda$ regime. 
In the case of higher-point functions, we are interested in general correlation functions of the form
\begin{equation}
\label{enecorr}
\langle \phi(x_1)^{n_1}\, \cdots \phi(x_r)^{n_r}\,\bar\phi(y_1)^{m_1}\, \cdots \bar \phi(y_s)^{m_s}\rangle\,\ ,\qquad \sum_{i=1}^r n_i =\sum_{j=1}^s m_j\ .
\end{equation}
We will assume the following scaling
\begin{equation}
    n_i = a_i n\ ,\qquad m_j=b_j n\ ,\qquad g\to 0,\ \ n\to\infty\ ,\ \ \ {gn^2}={\rm fixed}\ ,
\nonumber
\end{equation}
and fixed $a_i,\ b_j$.
In the case of the two-point function $\langle \phi(x)^{n} \bar \phi(y)^{n}\rangle$, it was shown in \cite{Arias-Tamargo:2019xld} that in the
double-scaling limit all higher loop diagrams vanish except those with a particular topology (the ``Kermit the frog" $L$-loop diagram), corresponding
to the case where two lines of each of the $L$ vertices join two of the $n$ lines of the operator  $\phi^n$ and the other two lines join two of the $n$ lines of the operator  $\bar \phi^n$. In particular, Feynman diagrams  having lines joining one vertex to another one vanish in the double-scaling limit. As a result, the two-point function can be exactly  computed by a complete resummation
of the surviving $L$-loop Feynman diagrams.

Alternatively, the double-scaling limit can be understood from a saddle-point calculation. This can be easily generalized to the general correlation function \eqref{enecorr}.
We first introduce  the scaled scalar field
\begin{equation}
    \sigma =g^{\frac14}\, \phi\ ,\qquad    \bar \sigma =g^{\frac14}\, \bar \phi\ .
\end{equation}
The general correlation function  \eqref{enecorr} is then given by
\begin{equation}
 \langle \phi(x_1)^{n_1}\, \cdots \phi(x_r)^{n_r}\,\bar\phi(y_1)^{m_1}\, \cdots \bar \phi(y_s)^{m_s}\rangle = \frac{1}{g^{\frac{m}{2}} Z}\int D\sigma D\bar\sigma \ e^{- S}\ ,\qquad m\equiv \sum_{j=1}^s m_j\ ,
\nonumber
\end{equation}
where the Euclidean action, including  source terms,  is given by
\begin{equation}
\label{S}
S= S_{\rm free}+ S_{\rm int}
\end{equation}
\begin{eqnarray}
 S_{\rm free} &=& \int d^4 x\,\left(g^{-\frac12} \partial\bar \sigma\,\partial\sigma 
-\sum_i n_i \delta(x-x_i)\log \sigma -\sum_j m_j\delta(x-y_j)\log \bar\sigma  \right)
 \nonumber\\
&=&  \int d^4 x\, \left(g^{-\frac12} \partial\bar \sigma\,\partial\sigma -\log\sigma(x_1)^{n_1}\cdots\sigma(x_r)^{n_r}\bar\sigma(y_1)^{m_1}\cdots \bar\sigma(y_s)^{m_s}\right)\ ,
\label{freess}
\\    
S_{\rm int} &=& \int d^4 x\, \frac{1}{4}\,(\bar\sigma \sigma)^2\ .
\label{Sinnn}
\end{eqnarray}
The saddle-point equations are given by
\begin{equation}
    \partial^2 \sigma =- ng^{\frac12} \sum_j b_j \delta(x-y_j)\frac{1}{\bar \sigma}+\frac12 g^{\frac12}\bar\sigma\sigma^2\ ,\qquad
      \partial^2 \bar\sigma =- ng^{\frac12} \sum_i a_i \delta(x-x_i)\frac{1}{ \sigma}+\frac12 g^{\frac12}\bar\sigma^2\sigma\ .
\nonumber
\end{equation}
In the double-scaling $g\to 0$, $n\to\infty$, with fixed $\lambda= n^2 g$, the cubic term vanishes.
The equations simply become
\begin{equation}
 \label{sadda}
   \bar\sigma  \partial^2 \sigma = - \lambda^{\frac12}\sum_j b_j \delta(x-y_j)\ ,\qquad \sigma \partial^2 \bar \sigma = - \lambda^{\frac12} \sum_i a_i \delta(x-x_i)\ .
\end{equation}

\subsection{Extremal correlators}

We shall first consider a special class of correlation functions where the resulting expressions in the double-scaling limit are conspicuously simple. These are the ``extremal" correlators
\begin{equation}
\label{excorr}
\langle \phi(x_1)^{n_1}\, \cdots \phi(x_r)^{n_r}\,\bar\phi(y)^{m}\rangle\,\ ,\qquad \sum_{i=1}^r n_i =m\ .
\end{equation}
The name  ``extremal" correlators is borrowed from $\mathcal{N}=2$ supersymmetric gauge theories, where, as explained earlier, correlation functions of this form are special by virtue of supersymmetry.
In the present case, there is of course no supersymmetry. Yet, for extremal correlators of the form \eqref{excorr}, the double-scaling limit is specially simple, singling out the particular topologies generalizing the Kermit diagrams of \cite{Arias-Tamargo:2019xld} with the two lines of each vertex
being distributed among the $r$ different points.
The reason of the simplicity of this correlator is more transparent in the saddle-point calculation,
which   for this case admits a simple solution.
The solution to \eqref{sadda} is
\begin{equation}
    \sigma =  \frac{\lambda^{\frac12} b}{\bar\sigma_0 (y)} G(x-y)\ ,\qquad  \bar\sigma = \bar\sigma_0 (y)  \sum_{i=1}^r  \frac{a_i G(x-x_i)}{bG(x_i-y)}\ ,
\end{equation}
where $G(x)$ is the Green's function
\begin{equation}
    G(x) = \frac{1}{4\pi x^2}\ ,\qquad \partial^2 G(x) =-\delta(x)\ ,
\nonumber
\end{equation}
Note that the factor $\bar\sigma_0 (y)= \bar\sigma (y)$ cancels out in computing the action.
Substituting this solution into the free part of the action, we obtain
\begin{equation}
    S_{\rm free} =-\log\sigma(x_1)^{n_1}\cdots\sigma(x_r)^{n_r}\bar\sigma(y)^m +m\ .
\end{equation}
This gives 
\begin{equation}
 \langle \phi(x_1)^{n_1}\, \cdots \phi(x_r)^{n_r}\,\bar\phi(y)^{m}\rangle _{\rm free} = \frac{m^m e^{-m}}{(4\,\pi^2)^m}\,\prod_{i=1}^r
 \frac{1}{|x_i-y|^{2 n_i}}\, \ .
 \end{equation}
 The factor $m^m e^{-m}$ is the leading approximation for $m!$ (the Gaussian integration in the saddle-point approximation completes
 the standard form of the de Moivre-Stirling formula $m!\approx \sqrt{2\pi m}\, m^m e^{-m}$).
 
 Next, let us consider the interaction term. 
\begin{eqnarray}
   S_{\rm int}= \int d^4 x \frac{1}{4}\,(\bar\sigma \sigma)^2 &=& \frac{\lambda }{4}\int d^4 x \, 
      G(x-y)^2  \left( \sum_i  \frac{a_i G(x-x_i)}{G(x_i-y)}\right)^2
      \nonumber\\
      &=& \frac{\lambda }{4}\left(\sum_{i=1}^r a_i^2 I(x_i,y) +2\sum_{i<j}^r  a_i a_j I(x_i,x_j,y) \right)
        \ , \nonumber
  \end{eqnarray}
  where \footnote{Details on the calculation of these integrals can be found in \cite{Freedman:1991tk} (see also \cite{Arias-Tamargo:2019xld}).}
    \begin{equation}
    \label{Ixx}
     I(x_i,y)\equiv \frac{1}{G(x_i-y)^2}\int  d^4 x \, G(x-y)^2 G(x-x_i)^2 = \frac{1}{4\pi^2} \log(\mu|x_i-y|)
  \end{equation}
  \begin{eqnarray}
     I(x_i,x_j,y)&\equiv & \frac{1}{G(x_i-y)G(x_j-y)}\int  d^4 x\, G(x-y)^2 G(x-x_i)G(x-x_j) 
     \nonumber\\
     &=& \frac{1}{8\pi^2} \log\Big(\mu\, \frac{|x_i-y|\,|x_j-y|}{|x_i-x_j|}\Big)\,,
\label{Ixxy}  \end{eqnarray}  
being $\mu$ is a reference mass scale, which in what follows will be set to one (see comments in appendix \ref{scalesection}).

Combining the free and the interacting part, we finally obtain
\begin{equation}
\label{Fextrem}
\langle \phi(x_1)^{n_1}\, \cdots \phi(x_r)^{n_r}\,\bar{\phi}(y)^{m}\rangle=\frac{m!}{(4\pi^2)^m\prod_{i=1}^r |x_i-y|^{2(n_i+\frac{\lambda \,a_i b}{32\pi^2})}\,\prod_{i<j}^r\,|x_i-x_j|^{-\frac{\lambda \,a_i\,a_j}{16\,\pi^2}} } \,.
\end{equation}
We can now check that this structure is consistent with the expected structure dictated by conformal symmetry.
Consider first the particular case of the three-point function, that is, $r=2$.
With no loss of generality, we can set $y=0$.
The result can be written in the equivalent form
\begin{equation}
\label{trespoint}
\langle \phi(x_1)^{n_1} \phi(x_2)^{n_2}\,\bar{\phi}(0)^{m}\rangle=\frac{m!}{(4\pi^2)^m  |x_1|^{\Delta_1+ \bar \Delta-\Delta_2}
|x_2|^{\Delta_2+\bar \Delta-\Delta_1} |x_1-x_2|^{\Delta_1+ \Delta_2-\bar \Delta} } \, ,
\end{equation}
where $m=n_1+n_2$ and
\begin{equation}
    \Delta_1 =n_1+ \frac{\lambda a_1^2}{32\pi^2}\ ,\quad \Delta_2 =n_2+ \frac{\lambda a_2^2}{32\pi^2}\ ,\quad 
    \bar \Delta =(n_1+n_2)+ \frac{\lambda (a_1+a_2)^2}{32\pi^2}\ .
\nonumber
\end{equation}
Higher-point extremal correlators are given explicitly by the remarkably simple formula \eqref{Fextrem}. 
When $r\geq 3$, the exponents in the formula \eqref{Fextrem} can
no longer be expressed purely in terms of the dimensions $\{\Delta_i, \bar \Delta \}$ as in the three-point function \eqref{trespoint}.

\medskip

\begin{figure}
    \centering
\begin{tikzpicture}[scale=1.1]

%LEFT DIAGRAM

\draw[thick] (-1,1) .. controls (-0.3,1.5) and (0.5,0.8) ..  (1,0);
\draw[thick] (-1,-1) .. controls (-0.3,-1.5) and (0.5,-0.8) ..  (1,0);
\draw[thick] (-1,-1) .. controls (-0.3,-0.85) and (0.5,-0.5) ..  (1,0);
\draw[thick] (-1,-1) .. controls (-0.3,-0.2) and (0.5,-0.2) ..  (1,0);

\draw[thick] (-1,1) .. controls (-0.6,1) and (-0.3,0.75) ..  (0,0.5);
\draw[thick] (-1,1) .. controls (-0.6,0.5) and (-0.3,0.5) ..  (0,0.5);
\draw[thick] (0,0.5) .. controls (0.3,0.5) and (0.6,0.4) ..  (1,0);
\draw[thick] (0,0.5) .. controls (0.3,0.1) and (0.6,0.0) ..  (1,0);
\draw[thick,fill] (0,0.5) circle (1.8pt);

\draw[thick,fill,red] (-1,1) circle (2pt);
\draw[thick,fill,red] (-1,-1) circle (2pt);
\draw[thick,fill,red] (1,0) circle (2pt);
\node at (-1.5,1) {$x_i$};
\node at (-1.5,-1) {$x_j$};
\node at (1.5,0) {$y$};
\node at (-1.5,0.15) {$\vdots$};

%RIGHT DIAGRAM

\draw[thick] (-1+4+1,1) .. controls (-0.3+4+1,1.5) and (0.5+4+1,0.8) ..  (1+1+4,0);
\draw[thick] (-1+4+1,1) .. controls (-0.3+4+1,0.85) and (0.5+4+1,0.5) ..  (1+4+1,0);
\draw[thick] (-1+4+1,-1) .. controls (-0.3+4+1,-1.5) and (0.5+4+1,-0.8) ..  (1+4+1,0);
\draw[thick] (-1+4+1,-1) .. controls (-0.3+4+1,-0.85) and (0.5+4+1,-0.5) ..  (1+4+1,0);

\draw[thick] (3+1,1) .. controls (3.3+1,0.3) and (3.7+1,0.2) .. (4+1,0);
\draw[thick] (3+1,-1) .. controls (3.3+1,-0.3) and (3.7+1,-0.2) .. (4+1,0);
\draw[thick] (4+1,0) .. controls (4.3+1,0.3) and (4.7+1,0.2) .. (5+1,0);
\draw[thick] (4+1,0) .. controls (4.3+1,-0.3) and (4.7+1,-0.2) .. (5+1,0);
\draw[thick,fill] (4+1,0) circle (1.8pt);

\draw[thick,fill,red] (-1+4+1,1) circle (2pt);
\draw[thick,fill,red] (-1+4+1,-1) circle (2pt);
\draw[thick,fill,red] (1+4+1,0) circle (2pt);
\node at (-1.5+4+1,1) {$x_i$};
\node at (-1.5+4+1,-1) {$x_j$};
\node at (1.5+4+1,0) {$y$};
\node at (-1.5+4+1,0.15) {$\vdots$};

\end{tikzpicture}
\caption{Types of diagrams that contribute to the extremal correlators.}
\label{diagram}
\end{figure}
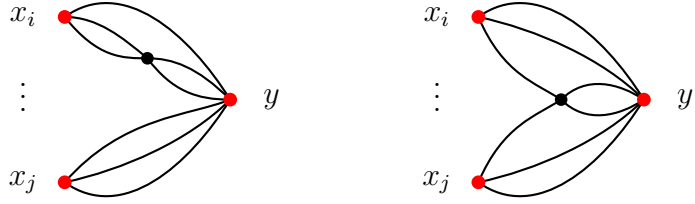

Summarizing, we found the exact ``extremal" correlators in the double-scaling limit where all charges go to infinity scaling in the same way. The result represents the resummation of the infinite number of $L$-loop Feynman diagrams that survive the limit.
These are shown in figure \ref{diagram} and generalize the ``Kermit the frog" diagrams described in detail in \cite{Arias-Tamargo:2019xld}.
The existence of the limit can be understood
from the saddle-point analysis, which led  to finite expressions
that become exact at $n=\infty $.
For large, but finite, charges, the double-scaling limit can be viewed as the leading result in a $1/n$ expansion.
The next $O(1/n)$ terms in the expansion may be systematically derived from corrections to the saddle point approximations,
obtained from the Taylor expansion of the action around the saddle-point.

\medskip

\subsection{Non-extremal correlators}

Let us now discuss general (``non-extremal") correlation functions. The general solution to \eqref{sadda}  is given by

\begin{equation}
\label{solsigma}
\sigma(x)=\lambda^{\frac{1}{2}}\sum_{j=1}^s\,\frac{b_j}{\bar{\sigma}(y_j)}\,G(x,\,y_j)\,,\qquad \bar{\sigma}(x)=\lambda^{\frac{1}{2}}\sum_{i=1}^r\,\frac{a_i}{\sigma(x_i)}\,G(x,\,x_i)\,,
\end{equation}
One can check that these equations are consistent provided 
$\sum_{i=1}^r n_i=\sum_{j=1}^s m_j$.  General correlation functions
can be obtained by substituting \eqref{solsigma} into the action
\eqref{freess}, \eqref{Sinnn}. In what follows we shall focus on the four-point function.

\subsubsection{Four-point non-extremal correlator}

As an explicit example, let us consider the  case
$r=s=2$, {\it i.e.} the four-point function
\begin{equation}
\langle \phi(x_1)^{n_1}\,\phi(x_2)^{n_2}\,\bar{\phi}(y_1)^{m_1}\,\bar{\phi}(y_2)^{m_2}\rangle\,,\qquad n_1+n_2=m_1+m_2\,.
\end{equation}
In addition, in this subsection we shall consider the particular case
\begin{equation}
a_1=a_2=b_1=b_2=1\,,
\end{equation}
so that $n_i=m_i=n$. Then

\begin{equation}
\label{solsig}
\sigma(x)=\sigma_0(x_2)\, \frac{ \sqrt{\frac{G(x_1,\,y_2)}{G(x_2,\,y_1)}}\,G(x,\,y_1)+\sqrt{\frac{G(x_1,\,y_1)}{G(x_2,\,y_2)}}\,G(x,\,y_2) }{\sqrt{G(x_1,\,y_2)\,G(x_2,\,y_1)}+\sqrt{G(x_1,\,y_1)\,G(x_2,\,y_2)}}\,\,,
\end{equation}
and

\begin{equation}
\label{solbar}
\bar{\sigma}(x)=\frac{\lambda^{\frac{1}{2}}}{\sigma_0(x_2)}\,\Big(G(x,\,x_2)+\sqrt{\frac{G(x_2,\,y_1)\,G(x_2,\,y_2)}{G(x_1,\,y_1)\,G(x_1,\,y_2)}}\,G(x,\,x_1)\Big)\,.
\end{equation}
The factor $\sigma_0(x_2)= \sigma(x_2) $  cancels out when computing the action. 
Substituting the solution into the free part of the action,
given in \eqref{freess}, we obtain

\begin{equation}
S_{\rm free}= 2n  -n \log \lambda \Big(\sqrt{G(x_1,\,y_2)\,G(x_2,\,y_1)}+\sqrt{G(x_1,\,y_1)\,G(x_2,\,y_2)}\Big)^{2}\,.
\end{equation}
It is convenient to rename $(y_1,y_2)\to(x_3,x_4)$ and define
$r_{ij}\equiv |x_i-x_j|$. Thus we obtain

\begin{equation}
\langle \phi(x_1)^{n}\,\phi(x_2)^{n}\,\bar{\phi}(x_3)^{n}\,\bar{\phi}(x_4)^{n}\rangle _{\rm free} =
\frac{n^{2n}e^{-2n}}{(4\pi^2 )^{2n}} \left( \frac{1}{r_{14}r_{23}}+\frac{1}{r_{13}r_{24}}
\right)^{2n}\ .
\label{libre}
\end{equation}

Let us now compute the interaction term.
Substituting the solutions \eqref{solsig}, \eqref{solbar}  for $\sigma$ and $\bar \sigma$ into \eqref{Sinnn}, we get an expression with nine integrals. Using the formulas \eqref{Ixx}, \eqref{Ixxy} and the integral computed in \cite{Dolan:2000uw} (see also \cite{Integral1,Integral2,Integral3})

\begin{equation}
    \int d^4x\, G(x,\,x_1)\,G(x,\,x_2)\,G(x,\,x_3)\,G(x,\,x_4)= \frac{H}{2^8\pi^6\,r_{13}^2\,r_{24}^2}\,,
\end{equation}
where

\begin{equation}
    H=\frac{1}{1-x-y}\Big( \log x(1-y) \, \log\frac{y}{1-x}-2\,{\rm Li}_2(x)+2\,{\rm Li}_2(1-y)\Big)\,,
\end{equation}
with

\begin{equation}
x=\frac{\rho\,u^2}{1+\rho\,u^2}\,,\quad y=\frac{\rho v^2}{1+\rho\,v^2}\,,\quad \rho=\frac{2}{1-u^2-v^2-\lambda}\,,\quad \lambda=\sqrt{(1-u^2-v^2)^2-4\,u^2\,v^2}\,,
\nonumber
\end{equation}
being $u$, $v$ the conformal ratios

\begin{equation}
u\equiv \frac{r_{12}r_{34}}{r_{13}r_{24}}\,,\qquad v \equiv \frac{r_{14}\,r_{23}}{r_{13}\,r_{24}}\, ;
\end{equation}
one finds that 

\begin{equation}
    \label{Sinte}
    S_{\text{int}}= \frac{\lambda}{16\pi^2}\,\log\frac{r_{13}r_{24}}{r_{12}r_{34}}+\frac{\lambda}{16\pi^2}\,\log\frac{r_{14}r_{23}}{r_{12}r_{34}}
    +\frac{\lambda}{16\pi^2}\,\log(r_{12}\,r_{34})+S'_{\text{int}}\,,
\end{equation}    
where
\begin{equation}
S'_{\text{int}}\equiv    \frac{\lambda}{16 \pi^2 }\frac{1}{(r_{14}r_{23}+r_{13} r_{24})^2}\bigg(
 H\,r_{14}^2\,r_{23}^2
-r_{13}^2r_{24}^2\log\frac{r_{13}r_{24}}{r_{12}r_{34}}-r_{14}^2r_{23}^2\log\frac{r_{14}r_{23}}{r_{12}r_{34}}  \bigg)\,.
\end{equation}

Thus, altogether, we obtain
\begin{equation}
   \langle \phi(x_1)^{n}\,\phi(x_2)^{n}\,\bar{\phi}(x_3)^{n}\,\bar{\phi}(x_4)^{n}\rangle=\frac{
   (n!)^2
}{(4\pi^2)^{2n}} \frac{(r_{14}r_{23}+r_{13} r_{24})^{2n}  \left( r_{12}r_{34}\right)^{\frac{\lambda}{16\pi^2}}}{\left(r_{14}r_{23}r_{13}r_{24}\right)^{2\Delta}}    \ e^{-S'_{\rm int}} \ .
\label{New4point}
\end{equation}

The final expression \eqref{New4point} has the symmetries under the exchanges $x_1\leftrightarrow x_2 $ and  $x_3\leftrightarrow x_4 $. These symmetries are 
 not manifest in the term with $H$, but they can be shown to hold using standard properties of ${\rm Li}_2(x)$
(see discussion in appendix C of  \cite{Dolan:2000uw}). The four-point function  \eqref{New4point} also has the expected singular behavior 
 in the channels $x_1=x_3$, $x_1=x_4$, $x_2=x_3$, $x_2=x_4$, with a power governed by the full scaling dimension $\Delta$ of the operators, including the anomalous dimension.
Here we have used the property  that $S'_{\rm int}$ is regular at any coinciding points, as can be shown using the above formula for $H$. 
 While the free part \eqref{libre} does not contain any singularity in the channels $r_{12}=0$ and $r_{34}=0$ because of charge conservation, due to the interaction there is a behavior 
 $\left( r_{12}r_{34}\right)^{\frac{\lambda}{16\pi^2}}$. This behavior was already present in the extremal  correlators. 
 The terms with $\log r_{12} $ and $\log r_{34}$ in  $S_{\text{int}}'$ exactly cancel out with similar terms originating from $H$ in the limit where either $r_{12}\to 0$ or $r_{34}\to 0$, so there is no extra
 contribution to this behavior.
 As a non-trivial check, we must recover the extremal three-point function \eqref{trespoint}
 in the limit $x_4\to x_3$. We obtain

 \begin{equation}
 \label{limitetres}
 \lim_{x_4 \to x_3} \langle \phi(x_1)^{n}\,\phi(x_2)^{n}\,\bar{\phi}(x_3)^{n}\,\bar{\phi}(x_4)^{n}\rangle =  \frac{(2n)!}{(4\pi^2)^{2n} }\, 
 \frac{ (\ell\, r_{12})^{\frac{\lambda}{16\pi^2}}}{ r_{13}^{\bar \Delta }
r_{23}^{\bar \Delta} } \ ,\qquad \bar\Delta =2n+\frac{\lambda}{8\pi^2}\ ,
 \end{equation}
where $\ell =r_{34}\to 0$. This reproduces \eqref{trespoint} for $n_1=n_2=n$, $m=2n$,
with an extra factor $\ell $ multiplying $r_{12}$. This factor is
to be absorbed into the reference scale $\mu$; see  discussion in appendix \ref{scalesection}.

\medskip

The important prediction of the double-scaling limit is that the
$O(\lambda)$ correction exponentiates. The saddle-point method exactly computes the full resummation of the surviving multiloop Feynman diagrams in the double-scaling limit.
The saddle point approximation receives $1/n$ corrections that we are not computing and reorganize into a more general expansion
in powers of $1/n$ and $\lambda$.

\subsubsection{Generating functional for the free part}

Here we shall compute general higher-points correlation functions for the free theory by computing the  generating functional.
This will also serve as a cross-check of the free ($\lambda=0$) part of the
previous results.
We consider the following correlation function:
\begin{equation}
    \langle \prod_{i=1}^r e^{\alpha_i \phi (x_i)}\, \prod_{j=1}^s e^{\beta_j \bar \phi (y_j)}
    \rangle  _{\rm free} .
\label{genfun}
\end{equation}
The desired (free) correlator \eqref{enecorr} is then obtained by  expanding the generating functional  in powers  of $\alpha_i $ and $\beta_j$ and isolating the term with the required powers $n_i, \ m_j$.
Including the source terms, the action is given by
\begin{equation}
S_{\rm free} =\int d^{4}x\,\left(\partial\bar \phi\,\partial\phi- \sum_i\alpha_i \delta(x-x_i) \phi - \sum_j\beta_j \delta(x-y_j) \bar\phi\right)\,.
\end{equation}
The functional integral is Gaussian and can be computed exactly,
with no need of taking any large charge limit, by solving the saddle-point equations. These are given by 
\begin{equation}
   \partial^2 \phi = - \sum_{j=1}^s \beta_j \delta(x-y_j)\ ,\qquad \partial^2 \bar \phi= -  \sum_{i=1}^r \alpha_i \delta(x-x_i)\ .
\end{equation}
The advantage of working with exponential operators is that the   equations have now the straightforward solutions
\begin{equation}
    \phi(x)= \sum_{j=1}^s\beta_j G(x-y_j)\ ,\qquad \bar \phi(x)= \sum_{i=1}^r\alpha_i G(x-x_i)\ .
\end{equation}
Substituting these solutions into the action we obtain
\begin{equation}
   S_{\rm free}=  -\sum_{j=1}^s\sum_{i=1}^r \alpha_i\beta_j G(x_i-y_j)\ .
\end{equation}

Using this formula, we reproduce the previous results for the free part of the {\it extremal} correlators in a straightforward way.

Let us now consider  non-extremal correlators. These have a more complicated structure involving several sums of terms, which originate from
many new possible contractions arising in Feynman diagrams.
As an example, here we consider the four-point correlation function
\begin{equation}
G_4\equiv \langle \phi(x_1)^{n_1}\phi(x_2)^{n_2}  \bar\phi(y_1)^{m_1}\bar\phi(y_2)^{ m_2} \rangle  _{\rm free}\ .
\end{equation}
We have
\begin{equation}
 \langle  e^{\alpha_1 \phi (x_1)} e^{\alpha_2 \phi (x_2)}\,  e^{\beta_1 \bar \phi (y_1)}e^{\beta_2 \bar \phi (y_2)}
    \rangle  _{\rm free}= e^{\alpha_1\beta_1 G(x_1-y_1)}e^{\alpha_2\beta_2 G(x_2-y_2)}e^{\alpha_1\beta_2 G(x_1-y_2)}e^{\alpha_2\beta_1 G(x_2-y_1)}\ .
\nonumber
\end{equation}
Expanding  in powers of $\alpha_i, \ \beta_j$ and isolating the terms with given powers $\alpha_1^{n_1} \alpha^{n_2}\beta_1^{m_1}\beta_2^{m_2}$,
we find
\begin{equation}
G_4 = n_1! n_2! m_1! m_2! \sum_k \frac{G(x_1-y_1)^k G(x_2-y_2)^{k+n_2-m_1}G(x_1-y_2)^{n_1-k} G(x_2-y_1)^{m_1-k} }{k!(n_1-k)!(k+n_2-m_1)!(m_1-k)!}    
\label{cuatropuntos}
\end{equation}
Thus far this is exact, valid for any values of $n_1,n_2,m_1,m_2$, with the sum over
$k$ restricted to $k\geq 0$, $k\leq m_1$, $k\geq m_1-n_2$, $k\leq n_1$.

Obtaining the correct asymptotic large charge behavior requires some care, as the approximation $(n-k)!\approx n!\,  n^{-k}$ cannot
be applied in (\ref{cuatropuntos}) because this holds for $k\ll n$ and terms with $k\sim n$ give a relevant contribution to the sum.
To illustrate this, let us consider in particular the case $n_1=n_2=m_1=m_2\equiv n$. Then we get
\begin{eqnarray}
G_4 &=& (n!)^4\sum_{k=0}^n \frac{G(x_1-y_1)^k G(x_2-y_2)^{k}G(x_1-y_2)^{n-k} G(x_2-y_1)^{n-k} }{k!^2(n-k)!^2} 
 \nonumber\\
 &=& \frac{ n!^2}{(4\pi^2)^{2n}}\frac{1}{r_{14}^{2n}\,r_{23}^{2n}}\ {}_2F_1(-n,-n,1,v^2)\,, 
 \label{finalex}
\end{eqnarray}
where we renamed $(y_1,\,y_2)\rightarrow (x_3,\,x_4)$. This formula is in agreement with the results presented in 
\cite{Dolan:2000ut} for the cases
$n=1$ and $n=2$,  given by
(6.17) and (6.21) in \cite{Dolan:2000ut} (for a real scalar field). Explicitly,

\begin{equation}
\frac{1}{r_{14}^{2n}\,r_{23}^{2n}}\, _2F_1(-n,\,-n,\,1,\,v^2)=
\begin{cases} u+\frac{u}{v}\quad &{\rm if }\ \  n=1\ ,  \\ 
u^2 +\frac{u^2}{v^2}+4\,\frac{u^2}{v}\quad &{\rm if }\ \  n=2 \ . \end{cases}
\end{equation}
The missing term ``1" in (6.2) of  \cite{Dolan:2000ut} is easily understood, as it comes from the identity operator which in the present $O(2)$ case cannot be exchanged in the $\phi(x_1)\,\phi(x_2)$ fusion due to charge conservation. As a further consistency check, 
in the limit $x_4\rightarrow x_3$ we find

\begin{equation}
\langle \phi(x_1)^{n}\,\phi(x_2)^{n}\,\bar{\phi}(x_3)^{2n}\rangle _{\rm free} = \frac{(2n)!}{(4\pi^2)^{2n}}\, \,\frac{1}{r_{13}^{2n}\,r_{23}^{2n}}\,,
\end{equation}
which is precisely the free part of the 3-point function (\textit{c.f} eq.\eqref{trespoint} for $\lambda=0$).

The exact result \eqref{finalex} can be used to cross-check the free part computed earlier in   \eqref{libre}. 
The asymptotic large $n$ behaviour can be obtained from the integral representation of the hypergeometric function, which at large $n$ is dominated by a saddle-point (see \textit{e.g.} \cite{hypergeometric}). This gives 
\begin{equation}
    {}_2F_1(-n,-n,1,v^2) \approx \frac{1}{\sqrt{4\pi n}} \frac{(1+v)^{1+2n}}{v^{\frac12 }}\ .
\nonumber
\end{equation}
Substituting this formula into \eqref{finalex}, we obtain

\begin{equation}
\langle \phi(x_1)^{n}\,\phi(x_2)^{n}\,\bar{\phi}(x_3)^{n}\,\bar{\phi}(x_4)^{n}\rangle_{\rm free} \approx \frac{ n!^2}{(4\pi^2)^{2n}}\frac{1}{\sqrt{4\pi n}}\frac{1}{ (r_{14}r_{23}r_{24}r_{13})^{n} }\ 
 \left( \sqrt{v}+\frac{1}{\sqrt{v}}\right)^{1+2n}\ .
\end{equation}
For large $n$, this coincides with  \eqref{libre}.

\section*{Acknowledgements}

We would like to thank A. Kehagias for useful insights and collaboration in an early stage of this work.
G.A-T and D.R-G are partially supported by the Spanish government grant MINECO-16-FPA2015-63667-P. They also acknowledge support from the Principado de Asturias through the grant FC-GRUPIN-IDI/2018/000174. G.A-T is supported by the Spanish government scholarship MCIU-19-FPU18/02221. J.G.R. acknowledges financial support from projects 2017-SGR-929, MINECO
grant FPA2016-76005-C.

\begin{appendix}

\section{On the scale dependence of correlation functions}\label{scalesection}

It is worth noting that the first two terms in \eqref{Sinte} and $S'_{\rm int}$ are dimensionless quantities (which can in fact be written in terms of the standard conformal ratios). 
Upon restoring the reference mass scale $\mu$, this appears only in the term $\log(\mu^2\,r_{12}\,r_{34})$ term. One may wonder how, in a CFT, a non-trivial dependence on a scale appeared in a correlation function. To understand this point, let us first consider the case of the two-point function written it in terms of dimensionless operators using the reference scale $\mu$. This leads to 

\begin{eqnarray}
    \langle \Big(\frac{\phi(x_1)}{\mu}\Big)^n\,\Big(\frac{\bar{\phi}(x_2)}{\mu}\Big)^n\rangle&=&\frac{n!}{(4\pi^2)^n}\,\frac{e^{-S_{\rm int}}}{(\mu\,|x_1-x_2|)^{2n}}=\frac{n!}{(4\pi^2)^n}\,\frac{e^{-\frac{\lambda}{16\pi^2}\,\log(\mu\,|x_1-x_2|)}}{(\mu\,|x_1-x_2|)^{2n}}\nonumber \\ &=&\frac{n!}{(4\pi^2)^n}\,\frac{1}{(\mu\,|x_1-x_2|)^{2\,\Delta}} \,,\qquad \Delta=n+\frac{\lambda}{32\,\pi^2}\,.
\end{eqnarray}
Thus, the $\mu$ dependence in the argument of the logarithm is precisely what it is required to soak up the dimensions of $x$ as it should be for a correlator of dimensionless operators. In other words, the $\mu$ dependence in the argument of the logarithm is reflecting the fact the operator has anomalous dimension. 

Now consider the four-point function  \eqref{New4point}. Similarly, the 
factor of $\mu$  arising from  the term $\log(\mu^2\,r_{12}\,r_{34})$ in \eqref{Sinte} combines with the factor $\mu^{-4n}$  to give a net factor $\mu^{-4\Delta}$, which is, in this case, the expected factor given that each operator has dimension $\Delta$. Restoring the  $\mu $ dependence,
the non-extremal four-point  function is given by
\begin{eqnarray}
 \frac{1}{\mu^{4n}}\langle \phi(x_1)^n \phi(x_2)^n \bar \phi(x_3)^n \bar\phi(x_4)^n \rangle
 &=& \frac{
   (n!)^2
}{(4\pi^2)^{2n}} \frac{\mu^{4n}(r_{14}r_{23}+r_{13} r_{24})^{2n}  \left( \mu^2 r_{12}r_{34}\right)^{\frac{\lambda}{16\pi^2}}}{\mu^{8\Delta} \left( r_{14}r_{23}r_{13}r_{24}\right)^{2\Delta}}    \ e^{-S'_{\rm int}} 
\nonumber\\
&=& \frac{
   (n!)^2
}{(4\pi^2)^{2n}} \frac{(r_{14}r_{23}+r_{13} r_{24})^{2n}  \left(  r_{12}r_{34}\right)^{\frac{\lambda}{16\pi^2}}}{\mu^{4\Delta} \left( r_{14}r_{23}r_{13}r_{24}\right)^{2\Delta}}    \ e^{-S'_{\rm int}} \ .
\end{eqnarray}
One can check that the same property holds for the general extremal correlator
\eqref{Fextrem}: the only $\mu $-dependence in 
$\mu^{-\sum_i n_i}\mu^{-m}\langle \phi(x_1)^{n_1}\, \cdots \phi(x_r)^{n_r}\,\bar{\phi}(y)^{m}\rangle$ is in a factor $\mu^{-\sum_i\Delta_i}\mu^{-\bar\Delta}$
on the RHS, as expected.

\end{appendix}

\end{document}